\documentclass[twocolumn,aps,prc,superscriptaddress,floatfix]{revtex4}
\usepackage{amssymb}
\usepackage{amsmath,bm}
\usepackage[normalem]{ulem}
\usepackage[dvips]{color}
\usepackage{booktabs}
\usepackage{graphicx}
\renewcommand{\sout}{\bgroup \color[rgb]{1,0,0}\ULdepth=-.5ex \ULset}

\begin{document}

\title{Light nuclei production as a probe of the QCD phase diagram}

\author{Kai-Jia Sun\footnote{%
sunkaijia@sjtu.edu.cn}}
\affiliation{School of Physics and Astronomy and Shanghai Key Laboratory for
Particle Physics and Cosmology, Shanghai Jiao Tong University, Shanghai 200240, China}
\author{Lie-Wen Chen\footnote{%
Corresponding author: lwchen$@$sjtu.edu.cn}}
\affiliation{School of Physics and Astronomy and Shanghai Key Laboratory for
Particle Physics and Cosmology, Shanghai Jiao Tong University, Shanghai 200240, China}
\author{Che Ming Ko\footnote{%
ko@comp.tamu.edu}}
\affiliation{Cyclotron Institute and Department of Physics and Astronomy, Texas A\&M University, College Station, Texas 77843, USA}
\author{Jie Pu\footnote{%
pujiephy@sjtu.edu.cn}}
\affiliation{School of Physics and Astronomy and Shanghai Key Laboratory for
Particle Physics and Cosmology, Shanghai Jiao Tong University, Shanghai 200240, China}
\author{Zhangbu Xu\footnote{%
xzb@bnl.gov}}
\affiliation{Brookhaven National Laboratory, Upton, New York 11973, USA}
\affiliation{School of Physics \& Key Laboratory of Particle Physics and Particle Irradiation (MOE), Shandong University, Jinan, Shandong 250100, China}

\date{\today}

\begin{abstract}
It is generally believed that the quark-hadron transition at small values of baryon chemical potentials $\mu_B$ is a crossover but changes to a first-order phase transition with an associated critical endpoint (CEP) as $\mu_B$ increases. Such a $\mu_B$-dependent quark-hadron transition is expected to result in a double-peak structure in the collision energy dependence of the baryon density fluctuation in heavy-ion collisions with one at lower energy due to the spinodal instability during the first-order phase transition and another at higher energy due to the critical fluctuations in the vicinity of the CEP. By analyzing the data on the $p$, d and $^3$H yields in central heavy-ion collisions within the coalescence model for light nuclei production, we find that the relative neutron density fluctuation $\Delta \rho_n=\langle(\delta \rho_n)^2\rangle/\langle \rho_n\rangle^2$ at kinetic freeze-out indeed displays a clear peak at  $\sqrt{s_{NN}}=8.8$~GeV and a possible strong re-enhancement at $\sqrt{s_{NN}}=4.86$~GeV. Our findings thus provide a strong support for the existence of a first-order phase transition at large $\mu_B$ and its critical endpoint at a smaller $\mu_B$ in the temperature versus baryon chemical potential plane of the QCD phase diagram.
\end{abstract}
\maketitle

\section{Introduction}

Understanding the phase diagram of strongly interacting matter is of fundamental importance in nuclear physics, astrophysics and cosmology. Lattice quantum chromodynamics (LQCD) calculations~\cite{Fod04} and various effective models~\cite{Asa89,Ste98,Hat03} have suggested that the transition between the quark-gluon plasma (QGP) and the hadronic matter is a smooth crossover at vanishing baryon chemical potential ($\mu_B$), but likely  changes to a first-order phase transition at large $\mu_B$, with an associated critical endpoint (CEP) or a tricritical endpoint~\cite{Din15}. In terrestrial labs, heavy-ion collisions provide a unique tool to study the structure of the QCD phase diagram~\cite{Shu09,Che09,Mun16,Luo17,Bug17}. In particular, to search for the CEP and locate the phase boundary  in the QCD phase diagram  is the main motivation for the heavy-ion collision experiments being carried out  in the Beam Energy Scan (BES) program at the Relativistic Heavy Ion collider (RHIC) as well as those planned at the future Facility for Antiproton and Ion Research (FAIR), the Nuclotron-based Ion Collider facility (NICA), and the SPS Heavy Ion and Neutrino Experiment (NA61/SHINE).

\begin{figure}[!b]
\includegraphics[scale=0.26]{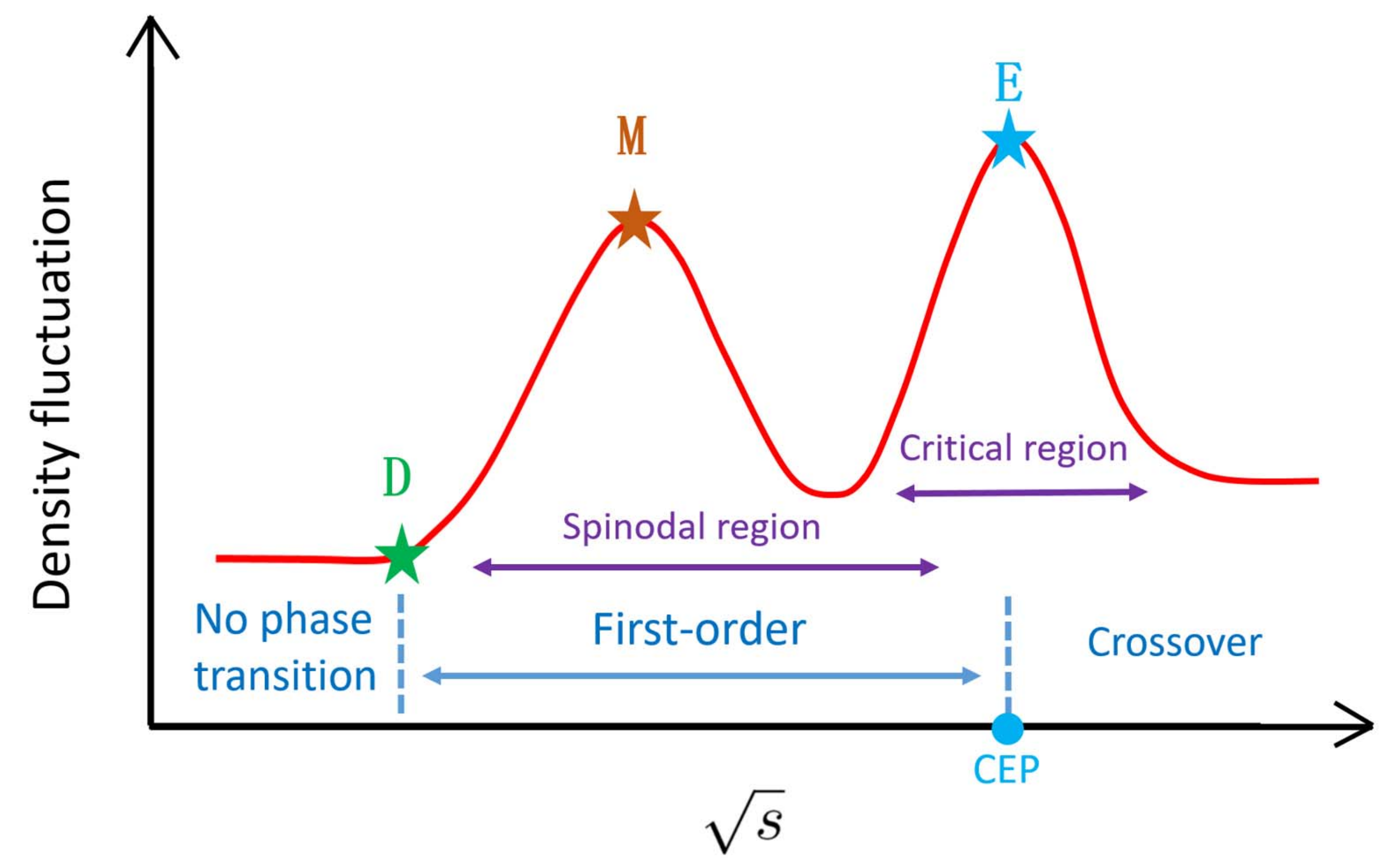}
\caption{Schematic depiction of the collision energy dependence of density fluctuations in heavy-ion collisions together with the corresponding phase regions in the QCD phase diagram. Point `D' indicates the beginning of first-order phase transition, `M' denotes the maximum caused by the spinodal instability and `E' denotes the maximum due to the CEP. }
\label{Fig1}
\end{figure}

In heavy-ion collisions, the created matter is expected to develop strong baryon density fluctuations when its evolution trajectory in the QCD phase diagram passes across the first-order phase transition line as a result of the spinodal instability~\cite{Ars07,Ran09,Stei12,Stei14,Hero14,Li16} or approaches the CEP due to a rapid increase of correlation length in the critical region~\cite{Ste98,Ber00}. In particular, for collisions at lower energies when the system enters the region of a first-order phase transition, the density fluctuation could reach a maximum at a collision energy $\sqrt{s}_\text{M}$ at which the system stays the longest time inside the unstable spinodal region, leading to the largest density inhomogeneity in the sytsem~\cite{Stei12}. With increasing collision energy, a second maximum in the density fluctuation would appear at a collision energy $\sqrt{s}_\text{E}$ at which the CEP is reached, leading to the onset of critical fluctuations~\cite{Ste98,Ber00}. This double-peak structure of the density fluctuation as a function of the collision energy $\sqrt{s}$ is schematically depicted in Fig.~\ref{Fig1}, in which the corresponding phase regions in the QCD phase diagram are also indicated.

Extracting the density fluctuations in heavy-ion collisions from experimental observables is a challenging task as only the particle momentum distributions are generally measured.  Because of the rapid expansion of the fireball formed in heavy-ion collisions, the enhanced density fluctuations caused by spinodal instability or critical fluctuation could, however, survive the final-state interactions and affect observables that are sensitive to nucleon density fluctuations and correlations at kinetic freeze-out when particles cease interacting. Besides the Hanbury-Brown-Twiss (HBT) interferometry of identical particles, which can provide information on the space-time structure of the particle emission source~\cite{Pra84,Ber88} and thus the effect of density fluctuations at hadronization~\cite{Che09}, these observables also include light nuclei production via nucleon coalescence~\cite{Sat81,Cse86,Dov91,ChenLW03,Bas16,KJSUN15,Zhu15,Mro17,KJSUN17,KJSUN17-1,Iva17}. Indeed, we have recently shown~\cite{KJSUN17,KJSUN17-1} that the yield ratio ${\mathcal O}_\text{p-d-t}=N_{\rm p}N_{\rm ^3H}/N_{\rm d}^2$ of produced proton ($N_{\rm p}$), deuteron~($N_{\rm d}$), and triton~($N_{\rm ^3H}$) depends on the relative neutron density fluctuation $\Delta \rho_n$, the neutron and proton density correlation $C_\text{np}$ as well as the phase-space volume~($V_\text{ph}$) occupied by nucleons at kinetic freeze-out. Assuming the ratio $\alpha = C_\text{np} / \Delta \rho_n$ to be independent of the collision energy, we have found~\cite{KJSUN17-1} from the measured yield ratio ${\mathcal O}_\text{p-d-t}$ in Pb+Pb collisions by the NA49 Collaboration~\cite{Ant16} at the CERN Super Proton Synchrotron (SPS) that the $\Delta \rho_n$ as a function of $\sqrt{s}$ shows a possible peak structure at $\sqrt{s}=8.8$ GeV, indicating that the density fluctuations become the largest at this energy.

In this Letter, we improve the above study by determining the collision energy dependence of $\Delta \rho_n$ and $C_\text{np}$ from that of $V_\text{ph}$ at kinetic freeze-out without assuming their ratio to be independent of the collision energy. Specifically, we argue that the phase-space volume $V_\text{ph}$ at kinetic freeze-out is related to the entropy per nucleon and can be determined from the temperature and volume at chemical freeze-out when chemical equilibrium is reached and the particles ratios are fixed. With the well determined chemical freeze-out temperature and volume from the statistical model fit to available experimental data, we can then simultaneously determine the collision energy dependence of $C_\text{np}$ and $\Delta \rho_n$. By analyzing the data in central collisions of Au+Au~\cite{Ahl99,Alb02} measured at the Brookhaven Alternating Gradient Synchrotron (AGS) and Pb+Pb measured by the NA49 Collaboration~\cite{Ant16} at SPS, we find that the $\Delta \rho_n$ displays a clear peak at $\sqrt{s_{NN}}=8.8$~GeV as found in Ref.~\cite{KJSUN17-1}
and a possible strong re-enhancement at $\sqrt{s_{NN}}=4.86$~GeV, confirming the double-peak structure shown in Fig.~\ref{Fig1}.

\section{Coalescence model for light nuclei production}

The coalescence model has been extensively and successfully used in studying light nuclei production in heavy-ion collisions~\cite{Sat81,Cse86,Dov91,ChenLW03,Bas16,KJSUN15,Zhu15,Mro17,KJSUN17,KJSUN17-1,Iva17}.
For deuteron production from an emission source of protons and neutrons, its number in the coalescence model is calculated from the overlap of the proton and neutron phase-space distribution functions $f_{p,n}({\bf x},{\bf k})$ with the Wigner function $W_{\rm d}({\bf x},{\bf k})$ of the deuteron internal wave function, i.e.,
\begin{eqnarray}
N_\text{d}&=&g_\text{d}\int \text{d}^3{\bf x}_1\int \text{d}^3{\bf k}_1\int \text{d}^3{\bf x}_2\int \text{d}^3{\bf k}_2 f_n({\bf x}_1,{\bf k}_1) \notag \\
&&f_p({\bf x}_2,{\bf k}_2)W_\text{d}({\bf x}_1-{\bf x}_2,({\bf k}_1-{\bf k}_2)/2), \label{Eq1}
\end{eqnarray}
with $g_d=3/4$ being the coalescence factor for deuteron. For protons and neutrons emitted from an isotropic and thermalized fireball of an effective temperature $T$ (after taking into account the flow effect) and volume $V$ and uniformly distributed in space, their distribution functions are then given by $f_{p,n}({\bf x},{\bf k})=\frac{2\xi_{p,n}}{(2\pi)^3}~e^{-\frac{k^2}{2mT}}$ where $m$ and $\xi_{p,n}$ are the nucleon mass and fugacities, respectively, and are normalized to their numbers $N_{p,n}=\int \text{d}^3{\bf x}\int \text{d}^3{\bf k}f_{p,n}({\bf x},{\bf k})=2\xi_{p,n} V\left(\frac{mT}{2\pi}\right)^{3/2}$.

Using the Gaussian or harmonic oscillator wave functions for the internal wave function of deuteron, as usually assumed in the coalescence model for deuteron production, its Wigner function is $W_\text{d}({\bf x},{\bf k})=8~e^{-\frac{x^2}{\sigma^2}}~e^{-\sigma^2 k^2}$  and is normalized according to $\int \text{d}^3{\bf x}\int \text{d}^3{\bf k}~W_d({\bf x},{\bf k})=(2\pi)^3$. With the coordinate and momentum transformations ${\bf X}=\frac{{\bf x}_1+{\bf x}_2}{2},~{\bf x}={\bf x}_1-{\bf x}_2,~{\bf K}={\bf k}_1+{\bf k}_2,~{\bf k}=\frac{{\bf k}_1-{\bf k}_2}{2}$, which are slightly different from those used in Ref.~\cite{KJSUN17}, the integrals in Eq.~(\ref{Eq1}) can be straightforwardly evaluated, leading to
\begin{eqnarray}
N_\text{d}&=&\frac{32 g_\text{d}\xi_n\xi_p}{(2\pi)^6}\int \text{d}^3{\bf X}\int \text{d}^3{\bf x}~e^{-\frac{x^2}{\sigma^2}}
\int \text{d}^3{\bf K}~e^{-\frac{K^2}{4mT}}    \notag \\
 &&\int \text{d}^3{\bf k}~e^{-k^2(\sigma^2+\frac{1}{mT})}\nonumber\\
&=&\frac{3}{2^{1/2}}\left(\frac{2\pi}{mT}\right)^{3/2}\frac{1}{\left(1+\frac{1}{mT\sigma^2}\right)^{3/2}}\frac{N_nN_p}{V}.  \label{Eq4}
\end{eqnarray}
The parameter $\sigma$ in Eq. (\ref{Eq4}) is related the root-mean-square radius $r_{\text d}$ of deuteron by $\sigma  = \sqrt{8/3}~r_{\text d}\approx 3.2$ fm, which is much smaller than the size of the fireball at kinetic freeze-out in central collisions of Pb+Pb at SPS energies and Au+Au at AGS energies considered here.

In relativistic heavy-ion collisions, the kinetic freeze-out temperature $T$ is typically about $100$ MeV before taking into consideration of the flow effect, we therefore have $mT\gg 1/\sigma^2$, and the number of produced deuteron can be approximated as
\begin{eqnarray}
N_\text{d}&\approx&\frac{3}{2^{1/2}}\left(\frac{2\pi}{mT}\right)^{3/2}\frac{N_nN_p}{V}. \label{Eq5}
\end{eqnarray}
We note that the above expression is exactly the same as that in Ref.~\cite{KJSUN17-1} obtained by assuming the Bjorken boost invariance for the expanding fireball.

For $^3{\rm H}$ production from the coalescence of a proton and two neutrons, a similar calculation gives its number as
\begin{eqnarray}
N_{^3\text{H}}&\approx&\frac{3^{3/2}}{4}\left(\frac{2\pi}{mT}\right)^{3}\frac{N_n^2N_p}{V^2}.\label{Eqh}
\end{eqnarray}

Eqs.~(\ref{Eq5}) and (\ref{Eqh}) can also be used to calculate the deuteron and triton rapidity densities $dN_{\rm d}/dy$ and $dN_{^3{\rm H}}/dy$ by replacing $N_n$, $N_p$ and $V$ with $dN_n/dy$, $dN_p/dy$ and $dV/dy$, respectively.

\section{Effects of density fluctuations and correlations}

For non-uniform nucleon density distributions, the neutron and proton fugacities $\xi_n$ and $\xi_p$ become coordinator dependent, and the factor $F=N_nN_p/V$ in Eq.~(\ref{Eq5}) should be replaced by
\begin{eqnarray}
F&=&\frac{1}{(\pi\sigma^2)^{\frac{3}{2}}}\int \text{d}^3{\bf x}_1\int \text{d}^3{\bf x}_2\rho_n({\bf x}_1)\rho_p({\bf x}_2)e^{-\frac{({\bf x}_1-{\bf x}_2)^2}{\sigma^2}} \label{AppEq1}\notag \\
&=&\frac{1}{(\pi\sigma^2)^{\frac{3}{2}}}\int \text{d}^3{\bf X}\int \text{d}^3{\bf x}\rho_n\left({\bf X}+\frac{{\bf x}}{2}\right)\rho_p\left({\bf X}-\frac{\bf x}{2}\right)\times\notag\\
&& e^{-\frac{{\bf x}^2}{\sigma^2}},
\end{eqnarray}
where $\rho_n({\bf x})$ and $\rho_p({\bf x})$ denote the neutron and proton density distributions in coordinate space, respectively. Through a gradient expansion of  $\rho_n({\bf x})$ and $\rho_p({\bf x})$ at point ${\bf X}$ to the first order, $F$ can be rewritten as
\begin{eqnarray}
F&\approx&\int \text{d}^3{\bf X}\rho_n({\bf X})\rho_p({\bf X})+\frac{1}{(\pi\sigma^2)^{\frac{3}{2}}}\int \text{d}^3{\bf X}\int \text{d}^3{\bf x}\nonumber\\
&&e^{-\frac{{\bf x}^2}{\sigma^2}}\left[\frac{{\bf x}}{2}\cdot\nabla \rho_n({\bf X})\right]\left[\frac{{\bf x}}{2}\cdot\nabla \rho_p({\bf X})\right].\label{AppEq3}
\end{eqnarray}
The term linear in ${\bf x}$ in the expansion vanishes because the integrand is an odd function. The value of the term quadratic in ${\bf x}$ depends on the nucleon density distribution in space. Assuming $\nabla \rho_n({\bf X})\sim \frac{\rho_n({\bf X})}{a}\bf{e_n}$ and $\nabla \rho_p({\bf X})\sim \frac{\rho_p({\bf X})}{a}\bf{e_p}$, where $\bf{e_n}$ and $\bf{e_p}$ are the unit vectors along the density gradients of the neutron and proton spacial distributions, respectively, and $a$ is the length over which their densities change appreciably, the second term in Eq.~(\ref{AppEq3}) can be approximated as
\begin{eqnarray}
F_2 \sim&&  \int \text{d}^3{\bf X} \rho_n({\bf X})\rho_p({\bf X})\times\notag\\
&&  \frac{1}{(\pi\sigma^2)^{3/2}}\int \text{d}^3{\bf x}e^{-\frac{{\bf x}^2}{\sigma^2}}\left[\frac{{\bf x\cdot \bf{e_n}}}{2a}\right]\left[\frac{{\bf x\cdot \bf{e_p}}}{2a}\right]. \label{AppEq4}
\end{eqnarray}
Because of $|{\bf x}\cdot {\bf e_n}|\le x$ and $|{\bf x}\cdot {\bf e_p}|\le x$, the second line of above equation is less than
\begin{eqnarray}
\frac{1}{(\pi\sigma^2)^{3/2}}\int \text{d}^3{\bf x}e^{-\frac{{\bf x}^2}{\sigma^2}}\left(\frac{x}{2a}\right)^2 = \frac{3}{8}\left(\frac{\sigma}{a}\right)^2. \label{AppEq5}
\end{eqnarray}
We thus have the following relation
\begin{eqnarray}
 |F_2|<\frac{3}{8}\left(\frac{\sigma}{a}\right)^2 \int \text{d}^3{\bf X} \rho_n({\bf X})\rho_p({\bf X}), \label{AppEq6}
\end{eqnarray}
which means that the magnitude of $F_2$ has an upper bound when ${\bf e_n}$ and ${\bf e_p}$ are always in the same direction or totally correlated.

The above analysis indicates that if the directions of ${\bf e_n}$ and ${\bf e_p}$ are not strongly correlated or $a$ is significantly larger than $\sigma$, then $F_2$ is much smaller than the first term in Eq.~(\ref{AppEq3}). Because of the randomness in the directions of ${\bf e_n}$ and ${\bf e_p}$ as well as the large spacial scale of the inhomogeneity in the nucleon density distribution at kinetic freeze-out, the second term in Eq.~(\ref{AppEq3}) can be neglected and the number of deuteron is then given by
\begin{eqnarray}
N_{\rm d} &\approx& \frac{3}{2^{1/2}}\left(\frac{2\pi}{mT}\right)^{3/2} \int \text{d}^3{\bf x} ~\rho_n({\bf x})\rho_p({\bf x}). \label{Eq6}
\end{eqnarray}
Similarly, the $^3$H yield is approximately given by
\begin{eqnarray}
N_{\rm ^3H} &\approx& \frac{3^{3/2}}{4}\left(\frac{2\pi}{mT}\right)^{3} \int \text{d}^3{\bf x} ~\rho_n^2({\bf x})\rho_p({\bf x}).  \label{Eq7}
\end{eqnarray}

To account for the density fluctuations, the neutron and proton densities in the emission source can be expressed as
\begin{eqnarray}
\rho_n({\bf x}) &=& \frac{1}{V}\int \rho_n({\bf x})\text{d}^3{\bf x}+\delta \rho_n({\bf x}) = \langle \rho_n\rangle+\delta \rho_n({\bf x}), \label{Eq8}\\
\rho_p({\bf x}) &=& \frac{1}{V}\int \rho_p({\bf x})\text{d}^3{\bf x}+\delta \rho_p({\bf x}) = \langle \rho_p\rangle+\delta \rho_p({\bf x}), \label{Eq9}
\end{eqnarray}
where $\langle \cdot\rangle = \frac{1}{V}\int \text{d}^3{\bf x}$ denotes the average  over space and $\delta \rho_n({\bf x})$~($\delta \rho_p({\bf x})$) with $\langle\delta \rho_n\rangle=0$~($\langle\delta \rho_p\rangle=0$) denotes the  neutron~(proton) density fluctuation from its average value $\langle \rho_n\rangle$~($\langle \rho_p\rangle$). Eq.~(\ref{Eq6}) can then be rewritten as
\begin{eqnarray}
N_{\rm d} &\approx&\frac{3}{2^{1/2}}\left(\frac{2\pi}{mT}\right)^{3/2} ~N_p\langle \rho_n\rangle (1+C_\text{np}), \label{Eq10}
\end{eqnarray}
where $C_\text{np} = \langle\delta \rho_n \delta \rho_p\rangle/(\langle \rho_n\rangle\langle \rho_p\rangle)$  characterizes the neutron and proton density correlation. For Eq.~(\ref{Eq7}), it can be approximately written as
\begin{eqnarray}
N_{\rm ^3H} &\approx&  \frac{3^{3/2}}{4}\left(\frac{2\pi}{mT}\right)^{3}N_p\langle \rho_n\rangle^2 (1+\Delta \rho_n+2C_\text{np}),\label{Eq11}
\end{eqnarray}
with $\Delta \rho_n=\langle (\delta \rho_n)^2\rangle /\langle \rho_n\rangle^2$ describing the relative neutron density fluctuation if we neglect the term $C_\text{nnp}=\langle (\delta \rho_n)^2\delta \rho_p\rangle/(\langle \rho_n \rangle^2\langle \rho_p \rangle)$.  Since $(\delta \rho_n)^2\delta \rho_p$  can have positive and negative values at different regions  of the emission source, there is a large cancellation when calculating its average value.  As a result, the magnitude of $C_\text{nnp}$  is much smaller than $\langle |(\delta \rho_n)^2\delta \rho_p|\rangle/(\langle \rho_n \rangle^2\langle \rho_p \rangle)$, which is smaller than $\langle|(\delta \rho_n)^3|\rangle/(\langle \rho_n \rangle^3)\sim (\Delta \rho_n)^{3/2}$  because of the non-perfect correlation between neutron and proton densities.  Since the value of $\Delta \rho_n$ is  less than one as we will see later,  the magnitude of  $C_\text{nnp}$ is much smaller than $\Delta \rho_n$ and thus can be safely neglected in obtaining Eq.~(\ref{Eq11}).

We would like to emphasize that both $C_\text{np}$ and $\Delta \rho_n$ are defined to be dimensionless to eliminate the effects due to the collision energy dependence of average neutron and proton densities. It is seen from Eqs.~(\ref{Eq10}) and (\ref{Eq11}) that the deuteron yield depends on $C_\text{np}$ but not $\Delta \rho_n$, while the triton yield depends on both. In addition, one sees from Eqs.~(\ref{Eq10}) and (\ref{Eq11}) that $C_\text{np}$ and $\Delta \rho$ can be uniquely determined from
\begin{eqnarray}
C_\text{np} &\approx& g_\text{p-d}R_\text{np} V_\text{ph} \mathcal{O}_\text{p-d} - 1, \label{Eq12} \\
\Delta \rho_n &\approx& g_\text{p-d-t}(1+C_\text{np})^2\mathcal{O}_\text{p-d-t}-2C_\text{np}-1,\label{Eq13}
\end{eqnarray}
with $g_\text{p-d}=\frac{2^{1/2}}{3(2\pi)^3}\approx 0.0019$, $g_\text{p-d-t}=9/4\times(4/3)^{3/2}\approx 3.5$, $\mathcal{O}_\text{p-d} = N_{\text{d}}/{N_\text{p}^2}$, ${\mathcal O}_\text{p-d-t}=N_{\rm p}N_{\rm ^3H}/N_{\rm d}^2$, $R_\text{np} = N_p/N_n = \langle \rho_p\rangle/\langle \rho_n\rangle$, and $V_\text{ph}=(2\pi mT)^{3/2}V$. Since neutrons are usually not measured in high energy heavy-ion collisions experiments, except in Ref.~\cite{Arm00}, we estimate in this work the ratio $R_\text{np}$ from the measured pion yield ratio by using the relation $N_p/N_n =(\pi^+/\pi^-)^{1/2}$ from the statistical model.

The quantity $V_\text{ph}=(2\pi mT)^{3/2}V$ in Eq.~(\ref{Eq12}) is the effective phase-space volume occupied by nucleons in the fireball at kinetic freeze-out~\cite{KJSUN17} and is directly related to the entropy per nucleon~($S/N$), which is given by the Sackur-Tetrode equation~\cite{Gri13} $S/N=5/2+\text{ln}(V_\text{ph}/N)$  in the non-relativistic Boltzmann approximation. If we assume $T^{3/2}V = \lambda T_\text{ch}^{3/2}V_\text{ch}$ with $T_\text{ch}$ and $V_\text{ch}$ being the chemical freeze-out temperature and volume, respectively, then the value of $\lambda$ is one for a gas of constant number of nucleons expanding isentropically after chemical freeze-out.  A recent microscopic transport model  study~\cite{Xu17} has shown, however,  that it is the entropy per particle that remains a constant after chemical freeze-out in heavy-ion collisions, which is dominated by pion production if the collision energy is high, and that all particle ratios remain essentially unchanged from the chemical to the kinetic freeze-out in these collisions, as assumed in the statistical model for particle production.  Based on the time evolution of $T$ and $V$ after chemical freeze-out  from this study~\cite{Xu17}, we find that the value of $\lambda$ ranges from $1.5$ at $7.7$~GeV to $1.7$ at $200$~GeV, suggesting that although $T^{3/2}V$ increases as fireball expands, $\lambda$ has a very weak dependence  on the collision  energy. Therefore, we can uniquely determine the values of $C_\text{np}$ and $\Delta \rho_n$ by using a constant $\lambda=1.6$.

It should be mentioned that light nuclei, such as d and $^3$H, are formed from nucleons in a very restricted phase-space volume of $\Delta x\sim$2 fm and $\Delta p\sim$0.1 GeV, and their production in heavy-ion collisions can thus be used as an ideal probe of the local nucleon density fluctuations at scales $\gtrsim$ 2 fm.  In contrast to studies~\cite{Luo17,Koc10,Ste09,Asa16} that mainly focus on the event-by-event fluctuations of conserved charges within a specific window in momentum space, the $\Delta \rho_n$ in Eq.~(\ref{Eq13}) has the advantage that it directly measures the spacial density fluctuation. To relate fluctuations in momentum space to those in coordinate space is highly non-trivial, especially at SPS and AGS energies where the longitudinal boost invariance is not well satisfied.

With the information on $C_\text{np}$ and $\Delta \rho_n$, we can further investigate the fluctuation of neutron and proton density difference~($\langle( \delta \rho_n-\delta \rho_p )^2\rangle$), which is closely related to the isospin density fluctuation since neutron and proton have opposite isospin quantum numbers. While the fluctuations in baryonic (B), electric (Q) and strange (S) charges have been extensively investigated~\cite{Luo17,Koc10}, the fluctuation in isospin is rarely studied in heavy-ion collisions. Recently it has been shown~\cite{Liu17} that the isospin effect at BES energies can dramatically change the critical behavior of baryon and charge number fluctuations, suggesting that the isospin effects could be strong in heavy-ion collisions at SPS and AGS energies. Defining the isospin density fluctuation $\Delta \rho_I$ as
\begin{eqnarray}
\Delta \rho_I = \frac{\langle( \delta \rho_n-\delta \rho_p )^2\rangle}{(\langle \rho_n\rangle+\langle \rho_p\rangle)^2} = \frac{R_\text{np}^2\Delta \rho_p-2R_\text{np}C_\text{np}+\Delta \rho_n}{(1+R_\text{np})^2}, \label{Eq14}
\end{eqnarray}
one sees that a negative $C_\text{np}$ can increase $\Delta \rho_I$. Although the isospin in high energy heavy-ion collisions is mostly carried by pions, the $\Delta \rho_I$ defined in terms of nucleons may still carry important information on the isospin density fluctuation as a result of the frequent interactions between pions and nucleons.

\section{Results and discussions}

\begin{table*}
\caption{Yields $dN/dy$ of $p$, d and $^3$H at midrapidity, together with the yield ratio $\pi^+/\pi^-$ measured in central Pb+Pb collisions at $20$ AGeV ($0-7\%$ centrality, $\sqrt{s_{NN}}=6.3$~GeV), $30$ AGeV~($0-7\%$ centrality, $\sqrt{s_{NN}}=7.6$~GeV), $40$ AGeV ($0-7\%$ centrality, $\sqrt{s_{NN}}=8.8$~GeV), $80$ AGeV ($0-7\%$ centrality, $\sqrt{s_{NN}}=12.3$~GeV), and $158$ AGeV ($0-12\%$ centrality, $\sqrt{s_{NN}}=17.3$~GeV) by the NA49 Collaboration~\cite{Ant16,Alt06,Alt08}. Also given are the chemical freeze-out temperature~$T_\text{ch}$~(GeV) and volume $V_\text{ch}$~(fm$^3$), the derived yield ratios $\mathcal{O}_{\text{p-d}}$  and $\mathcal{O}_{\text{p-d-t}}$, and the extracted $C_\text{np}$, $\Delta \protect\rho_n$ and $\Delta \protect\rho_I$. In obtaining $\mathcal{O}_{\text{p-d}}$ and $\mathcal{O}_{\text{p-d-t}}$, the weak decay contributions to the yield of proton from hyperons are corrected by using results from the statistical model~(see text for details).}
\begin{tabular}{c|c|c|c|c|c|c|c|c|c|c|c}
        \hline \hline
          $\sqrt{s_{NN}}$  &  $p$ & d & $^3$H($10^{-3}$)& $\pi^+/\pi^-$ &  $T_\text{ch}$ &$V_\text{ch}$ & $\mathcal{O}_{\text{p-d}}(10^{-4})$& $\mathcal{O}_{\text{p-d-t}}$ &$C_\text{np}$ & $\Delta \rho_n$ & $\Delta \rho_I$ \\
         \hline
           6.3&    46.1$\pm$2.1 &  2.094$\pm$0.168& $43.7(\pm 6.4)$& 0.86 &  0.131& 1389&10.5$\pm$0.11 &0.444$\pm$0.014 &-0.636$\pm$0.004& 0.475$\pm$0.007& 0.556$\pm$0.004  \\
           7.6&    42.1$\pm$2.0 & 1.379$\pm$0.111&  $22.3(\pm 3.4)$& 0.88& 0.139& 1212& 8.78$\pm$0.13& 0.465$\pm$0.019 &-0.707$\pm$0.004&0.551$\pm$0.007 & 0.629$\pm$0.004  \\
           8.8&     41.3$\pm$1.1 & 1.065$\pm$0.086&  $14.8(\pm 2.6)$&0.90&0.144& 1166&7.32$\pm$0.20   & 0.500$\pm$0.020 & -0.749$\pm$0.007& 0.606$\pm$0.045 & 0.677$\pm$0.006\\
           12.3&    30.1$\pm$1.0 &0.543$\pm$0.044 &$4.49(\pm 0.94)$&0.91&0.153& 1231& 7.70$\pm$0.11&  0.404$\pm$0.034&-0.693$\pm$0.004 &0.518$\pm$0.012 & 0.605$\pm$0.006	\\
           17.3&   23.9$\pm$1.0 & 0.279$\pm$0.023& $1.58(\pm 0.31)$&0.93&0.159& 1389& 6.66$\pm$0.01& 0.415$\pm$0.032 &-0.681$\pm$0.0004&0.507$\pm$0.011 & 0.594$\pm$0.006\\
        \hline  \hline
\end{tabular}
\label{tab1}
\end{table*}

Before presenting our results, we would like to emphasize that the protons from long-lived weak decays of hyperons should be excluded in the coalescence model calculations for d and $^3$H production because
they would appear outside the fireball. Summarized in Table~\ref{tab1} are the yields $dN/dy$ of $p$, d and $^3$H at midrapidity, together with the pion yield ratio $\pi^+/\pi^-$ measured in central Pb+Pb collisions at $20$ AGeV ($0-7\%$ centrality), $30$ AGeV~($0-7\%$ centrality), $40$ AGeV ($0-7\%$ centrality), $80$ AGeV ($0-7\%$ centrality), and $158$ AGeV ($0-12\%$ centrality) by the NA49 Collaboration~\cite{Ant16,Alt06,Alt08}. The proton yields in Table~\ref{tab1} are the corrected results after subtracting the contribution from weak decays of hyperons, which is taken to be 15\% of the total proton yield at all SPS energies~\cite{Alt06}. In the present study, we use instead the collision energy dependent fraction from the statistical model calculations~\cite{And06}, namely, 17.7\%, 20.0\%, 21.5\%, 25.0\% and 27.2\% at $\sqrt{s_{NN}}=$6.3, 7.6, 8.8, 12.3 and 17.3 GeV, respectively, to correct the proton yield. The yield ratios $\mathcal{O}_\text{p-d}$ and $\mathcal{O}_\text{p-d-t}$ shown in Table~\ref{tab1} are those based on this corrected proton yield, and their errors are estimated by assuming they are dominated by correlated systematic errors as a result of similar detector acceptance and phase-space extrapolation. The chemical freeze-out temperature $T_\text{ch}$ is calculated from the parametrization given in Ref.~\cite{Cle06} while $V_\text{ch}$ is obtained from Ref.~\cite{And14}. The neutron and proton density correlation~($C_\text{np}$) and the relative neutron density fluctuation~($\Delta\rho_n$) as well as $\Delta \rho_I$ are then calculated from the Eqs.~(\ref{Eq12})-(\ref{Eq14}). In calculating $\Delta \rho_I$, we have taken the relative proton density fluctuation $\Delta \rho_p$ to be the same as that of neutrons as expected from the isospin invariance of strong interaction. It is seen from Table \ref{tab1} that all $C_\text{np}$, $\Delta\rho_n$ and $\Delta \rho_I$ show significant non-monotonic energy dependence.

\begin{figure}
\includegraphics[scale=0.31]{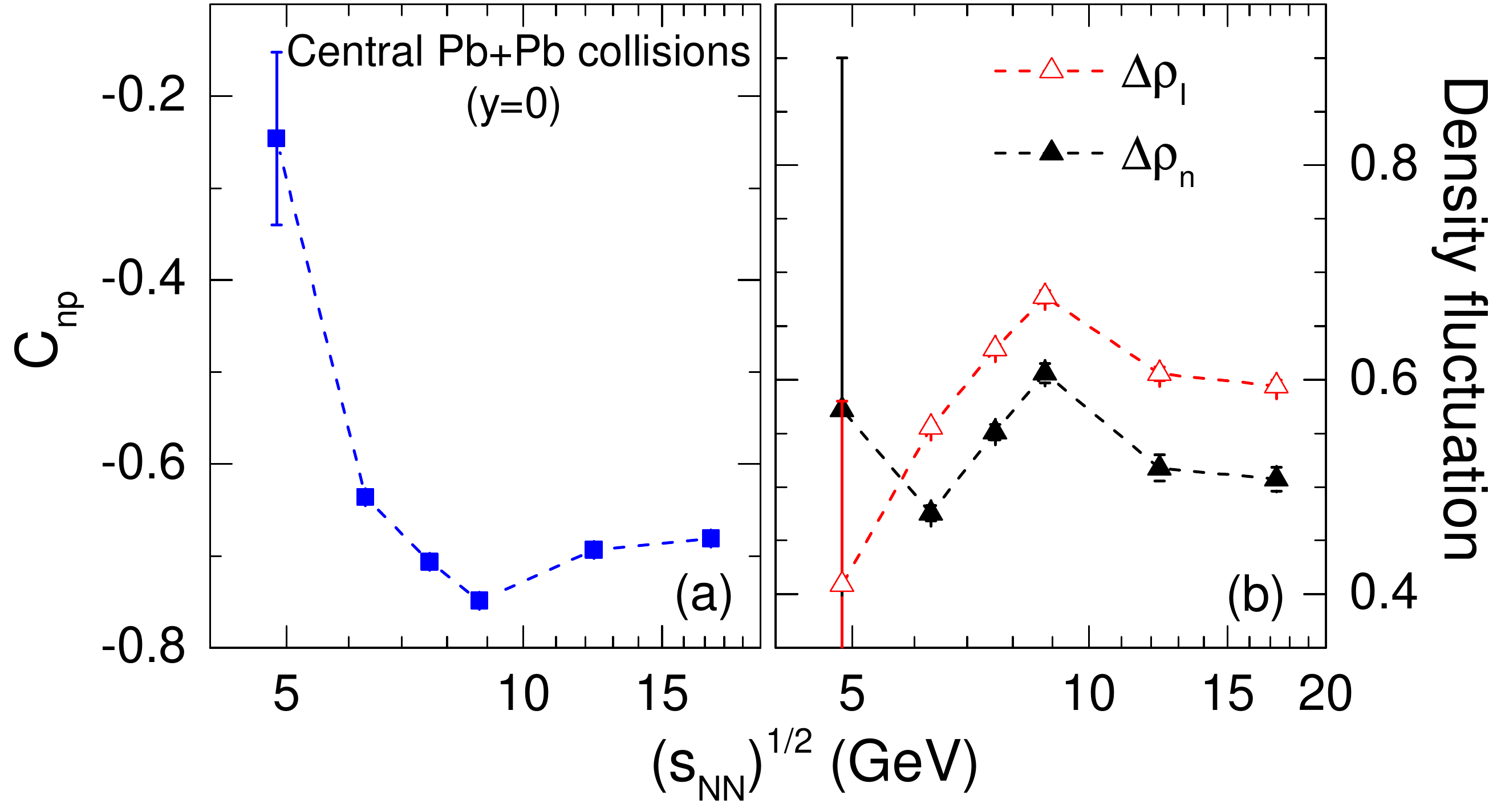}
\caption{Collision energy dependence of the neutron and proton density correlation $C_\text{np}$ (a) and the neutron and isospin density fluctuations $\Delta \protect\rho_n$ and $\Delta \protect\rho_I$ (b) in central Pb+Pb collisions at SPS energies and Au+Au collisions at AGS energies.}
\label{Fig2}
\end{figure}

To see more clearly their non-monotonic behaviors, we plot the collision energy dependence of $C_\text{np}$, $\Delta \rho_n$ and $\Delta \rho_I$   in Fig.~\ref{Fig2}.  We first focus on the results at SPS energies.  It is seen from Fig.~\ref{Fig2}~(a) that the neutron-proton density correlation $C_\text{np}$ has a non-monotonic behavior with a valley located at $\sqrt{s_{NN}}=$8.8 GeV. Besides, the extracted $C_\text{np}$ at SPS energies are all negative, indicating a strong negative correlation between the neutron and proton densities. From Fig.~\ref{Fig2}~(b), one sees that the neutron and isospin density fluctuations $\Delta \rho_n$ and $\Delta \rho_I$ show very similar non-monotonic behaviors with peaks also located at $\sqrt{s_{NN}}=8.8$ GeV. The obtained peak structure of $\Delta \rho_n$ is similar to that obtained in Ref.~\cite{KJSUN17-1}, where the ratio $\alpha=C_\text{np}/\Delta \rho_n$ is fixed at some constant values (e.g., $\alpha=$ $-$0.2, $-$0.1, 0, 0.1 and 0.2), but is much pronounced. From the extracted values of $\Delta \rho_n$ and $C_\text{np}$, we find that their ratio $\alpha$ has the values $-1.339 \pm 0.022$, $-1.282 \pm 0.018$, $-1.235 \pm 0.022$, $-1.339 \pm 0.032$ and $-1.342 \pm 0.029$ for $\sqrt{s_{NN}} = 6.3, 7.6, 8.8, 12.3~\rm and~17.3$ GeV, respectively. Although the value of $\alpha$ only has a weakly non-monotonic dependence on the collision energy, justifying its collision energy independence assumed in Ref.~\cite{KJSUN17-1}, its magnitude is significantly larger, leading thus to a much more pronounced and significant peak of $\Delta \rho_n$ at $\sqrt{s_{NN}} = 8.8 $ GeV than that found in Ref.~\cite{KJSUN17-1}.

The similar behavior of $\Delta \rho_n$ and $\Delta \rho_I$ with both peaked at $\sqrt{s_{NN}}=8.8$ GeV could be due to the same underlying physics of critical fluctuations in the vicinity of CEP. According to the universality of critical behavior, the singular parts of both $\Delta \rho_n$ and $\Delta \rho_I$ in the second-order phase transition scale with the correlation length $l$ as $l^{2-\eta}$, where $\eta$ denotes the critical exponent and is zero in the mean-field approximation, and diverge at the CEP in the QCD phase diagram. Due to the effects of critical slowing down~\cite{Ber00} and dynamical expansion in heavy-ion collisions, only modest but similar enhancements of $\Delta \rho_n$ and $\Delta \rho_I$ can be developed. As a result, the non-monotonic behaviors shown in Fig.~\ref{Fig2} are consistent with the scenario that the CEP is reached or closely approached in the produced QGP during its time evolution in central Pb+Pb collisions at around $\sqrt{s_{NN}}=8.8$ GeV. For lower (e.g., $\sqrt{s_{NN}}=6.3$ and 7.6 GeV) and higher (e.g., $\sqrt{s_{NN}}=12.3$ and 17.3 GeV) energies, both $\Delta \rho_n$ and $\Delta \rho_I$ would decrease because the correlation length $l$ quickly decreases as the evolution trajectory moves away from the CEP. From $\sqrt{s_{NN}}=6.3$ GeV to $8.8$ GeV, the correlation length $l$ effectively increases by about 13\% if we assume $\eta\ll 1$. This implies that the fourth-order baryon number cumulant $\langle(\delta N_B)^4\rangle_c~\sim ~l^{7-2\eta}$~\cite{Ste09} would increase by a factor of about $2.35$ from $\sqrt{s_{NN}}=6.3$ GeV to 8.8 GeV.  A similar enhancement of $l$ happens from $\sqrt{s_{NN}}=17.3$ GeV to $8.8$ GeV. It is very interesting to note the preliminary results that both the second-order and fourth-order cumulants of net proton distribution in central Au+Au collisions are also found to increase from $\sqrt{s_{NN}}=19.6$ GeV to 7.7 GeV~(see Fig. 8 in Ref. \cite{He17}).

From the parametrization in Ref.~\cite{Cle06} for the chemical freeze-out conditions based on the statistical model fit to available experimental data of hadron yields, the temperature at $\sqrt{s_{NN}}= 8.8$~GeV is estimated to be $T^\text{CEP} \sim 144$ MeV with a corresponding baryon chemical potential $\mu^\text{CEP}_B \sim 385$ MeV, which is close to  the predicted CEP from the LQCD~\cite{Fod04}, the Dyson-Schwinger equation~(DSE)~\cite{Xin14} and the hadronic bootstrap approach~\cite{Ant03}, but is much larger than that~($\sim 95$~MeV) inferred from a finite size scaling~(FSS) analysis of two-pion correlations~\cite{Lac15}. It is very likely that the critical region is reached within the energy region $\sqrt{s_{NN}}=6.3\sim17.3$~GeV. With the temperature and baryon chemical potential of about $131\sim159$ MeV and $481\sim229$ MeV~\cite{Cle06}, respectively, as determined from the statistical model, one can estimate the size of critical region to be $\Delta T/T^\text{CEP}\approx 0.1$ and $\Delta \mu_B\approx0.1$~GeV, which is consistent with the effective model calculations~\cite{Hat03}. The critical exponents, however, can not be determined from present data.

In the scenario depicted in Fig.~\ref{Fig1}, the effects from the first-order phase transition and from the CEP are related, because the CEP stays well on the top of the spinodal unstable region in the $T$-$\mu_B$ plane of the QCD phase diagram. Indeed, it has been estimated~\cite{Ran09} that the temperature~($T^\text{M}$) at point `M' is related to $T^\text{CEP}$ by $T^\text{M}/T^\text{CEP}\approx \frac{1}{3}-\frac{1}{2}$. From the parametrization in Ref.~\cite{Cle06} for the chemical freeze-out conditions, one can roughly estimate that $T^\text{M}\sim 50-70$~MeV and $\sqrt{s}_\text{M}\sim 2-3$~GeV, which is much smaller than the energies available at SPS, indicating that the effects from the first-order phase transition and from the CEP are well separated, making it possible to unambiguously identify them. In fact, according to Ref.~\cite{Stei12}, the enhanced density fluctuation due to spinodal instability happens at around $2$-$4$ AGeV~($\sqrt{s_{NN}}\sim 2.3-3$~GeV) and is in nice agreement with our estimate. Of course, the  definitive value of $\sqrt{s}_\text{M}$ should depend on the specific equation of state (EOS) adopted in the calculations, which is still largely unknown.

With decreasing collision energy to $\sqrt{s_{NN}}\sim 2-6$~GeV, the $\Delta \rho_n$ is  expected to rise again and reach a second maximum as a result of the spinodal instability as shown in Fig.~\ref{Fig1}. However, the collision energy dependence of $\Delta \rho_I$ and $\Delta \rho_n$ at $\sqrt{s_{NN}}\sim 2-6$~GeV might not coincide in this unstable spinodal region, due to the fact that the effect of spinodal instability in the isospin density is not as strong as that in the baryon density for heavy-ion collisions at AGS/SPS energies. To our best knowledge, the only existing data on the $p$, d, and $^3$H yields in the energy region $\sqrt{s_{NN}}\sim 2-6$~GeV are from central Au+Au collisions at $11.6$ AGeV/c~($\sqrt{s_{NN}}=4.86$ GeV), namely, $63.2\pm 1.7$ for proton~\cite{Ahl99}, $5.3\pm 0.6$ for deuteron~\cite{Ahl99} and $0.264\pm0.049$ for $^3$He~\cite{Alb02}. With $R_\text{np} \approx(0.84)^{1/2}=0.916$, $T_\text{ch}=0.117~$GeV~\cite{Cle06}, $V_\text{ch}=1980$ fm$^3$~\cite{And14} and using a constant 15\%~\cite{And06} weak decay contribution correction to the proton yield for simplicity, we obtain $C_\text{np}= -0.246 \pm 0.094$, $\Delta \rho_n=0.572 \pm 0.328$ and $\Delta \rho_I=0.409 \pm 0.171$, indicating a possible strong re-increase of $\Delta \rho_n$  when the collision energy is decreased from $\sqrt{s_{NN}}=6.3$~GeV  to $4.86$~GeV. In contrast, the $\Delta \rho_I$~($C_\text{np}$) continues to decrease~(increase) as clearly seen in Fig.~\ref{Fig2}.  However, the statistical uncertainties of $\Delta \rho_n$ and $\Delta \rho_I$ at $\sqrt{s_{NN}}=4.86$ GeV are  large, and more precise measurements are extremely important to confirm the present conclusion.

We would like to point out that some effects can potentially affect the above numerical results. These include the uncertainties in the value of $\lambda$ associated with the entropy of the expanding fireball as well as the chemical freeze-out temperature $T_\text{ch}$ and volume $V_\text{ch}$.  Also, the finite size of the fireball and the use of non-relativistic approximation for the nucleon momentum distributions~\cite{KJSUN17,KJSUN17-1} can affect the extracted values of $C_\text{np}$, $\Delta \rho_n$ and $\Delta \rho_I$. However, these effects on the above dimensionless quantities are expected to have a weak dependence on the collision energy, and the non-monotonic behaviors shown in Fig.~\ref{Fig2} should remain qualitatively similar.

\section{Conclusions and outlook}

We have proposed in the present study a double-peak structure in the collision energy dependence of the baryon density fluctuation in heavy-ion collisions as a probe to the structure of the QCD phase diagram, with the lower energy one due to the spinodal instability associated with a first-order quark-hadron phase transition and the higher energy one induced by the second-order phase transition at the CEP.  This double-peak structure seems to be supported by the collision energy dependence of the relative neutron density fluctuation $\Delta \rho_n=\langle(\delta \rho_n)^2\rangle/\langle \rho_n\rangle^2$ at kinetic freeze-out that we have extracted from analyzing the measured yields of $p$, d and $^3$H in central heavy-ion collisions at AGS and SPS energies within the coalescence model. In particular, we have found the $\Delta \rho_n$ to display a clear peak at $\sqrt{s_{NN}}=8.8$~GeV and a possible strong re-enhancement at $\sqrt{s_{NN}}=4.86$~GeV, suggesting that the CEP could have been reached or closely approached in central Pb+Pb collisions at $\sqrt{s_{NN}}=8.8$~GeV and the first-order phase transition could have occurred in central Au+Au collisions at $\sqrt{s_{NN}}=4.86$~GeV.

Although our results cannot tell whether the phase transition is due to deconfinement or the restoration of chiral symmetry, they provide a complementary evidence for the occurrence of a first-order phase transition and a critical endpoint in the QCD phase diagram to those that have been suggested in the literatures.  These include the irregularities of the trace anomaly $(\epsilon-3p)/T^4$ and quasi-plateaus of entropy per baryon ~\cite{Bug17} at two ranges of the center-of-mass collision energies of $\sqrt{s_{NN}}=3.8-4.9$~GeV~ and $\sqrt{s_{NN}}=7.6-9.2$~GeV, based on the non-smooth chemical freeze-out analysis~\cite{Bug13}; the explanation of the collision energy dependence of the $K^+/\pi^+$ ratio in terms of the chiral symmetry restoration~\cite{Pal16}; large higher-order moments in the event-by-event fluctuations of conserved charges~\cite{Luo17}; enhanced dilepton production~\cite{Li16,Wun16}; and the need of deconfinement in describing measured rapidity distribution~\cite{Iva17} and direct flow~\cite{Iva17,Bas16} of light nuclei.

To verify the present conclusion, it will be particularly important to carry out similar studies using microscopic transport model simulations as well as hydrodynamics calculations with the proper treatment of the equation of state and the critical fluctuations. Comparing results from these studies with future experimental data on light nuclei production at BES/RHIC, FAIR, NICA and NA61/SHINE will then allow for a more precise determination of the structure of the QCD phase diagram.

\section*{Acknowledgments}

The authors thank Vadim Kolesnikov and Peter Seyboth for providing the experimental data. This work was supported in part by the National Natural Science Foundation of China under Grant No. 11625521, the Major State Basic Research Development Program (973 Program) in China under Contract No. 2015CB856904, the Program for Professor of Special Appointment (Eastern Scholar) at Shanghai Institutions of Higher Learning, Key Laboratory for Particle Physics, Astrophysics and Cosmology,
Ministry of Education, China, the Science and Technology Commission of Shanghai Municipality (11DZ2260700), the US Department of Energy under Contract No. DE-SC0015266 and No. DE-SC0012704, as well as the Welch Foundation under Grant No. A-1358 and Shandong University.

\end{document}